\documentclass[aps,prd,twocolumn,showkeys,floats,showpacs, superscriptaddress,nofootinbib]{revtex4-1}
\pdfoutput=1 
\usepackage{graphicx}

\usepackage{amsmath,amssymb}

\usepackage{aas_macros}

\usepackage{natbib}
\usepackage[colorlinks=true]{hyperref}

\def\rhou{{\rho_{(0)}}}
\def\n0{{n_{(0)}}}
\def\T0{{T_{(0)}}}

\begin{document}
\title{Breaking of the equivalence principle in the electromagnetic sector and its cosmological signatures}

\author{Aur\'elien Hees}
\email{A.Hees@ru.ac.za}
\affiliation{Department of Mathematics, Rhodes University, Grahamstown 6140, South Africa}

\author{Olivier Minazzoli}
\email{ominazzoli@gmail.com}
\affiliation{Centre Scientifique de Monaco and UMR ARTEMIS, CNRS, University of Nice Sophia-Antipolis,
Observatoire de la C\^ote d'Azur, BP4229, 06304, Nice Cedex 4, France}

\author{Julien Larena}
\email{J.Larena@ru.ac.za}
\affiliation{Department of Mathematics, Rhodes University, Grahamstown 6140, South Africa}

%

\begin{abstract}
	This paper proposes a systematic study of cosmological signatures of modifications of gravity via the presence of a scalar field with a multiplicative coupling to the electromagnetic Lagrangian. We show that, in this framework, variations of the fine structure constant, violations of the distance-duality relation, evolution of the cosmic microwave background (CMB) temperature and CMB distortions are intimately and unequivocally linked. This enables one to put very stringent constraints on possible violations of the distance-duality relation, on the evolution of the CMB temperature and on admissible CMB distortions using current constraints on the fine structure constant. Alternatively, this offers interesting possibilities to test a wide range of theories of gravity by analysing several data sets concurrently. We discuss results obtained using current data as well as some forecasts for future data sets such as those coming from EUCLID or the SKA.
\end{abstract}

\maketitle

\section{Introduction}
The Einstein Equivalence Principle (EEP) is one of the building block of General Relativity (GR). This principle allows one to identify the effects of gravitation with space-time geometry. More precisely, it implies the existence of a space-time metric $g_{\mu\nu}$ to which matter is minimally coupled \cite{will:1993fk}. Mathematically, this implies that the action related to matter can be written as 
\begin{equation}\label{eq:matterAction}
 S_{\textrm{mat}}=\int d^4x\sqrt{-g} \mathcal L_{\textrm{mat}}(g_{\mu\nu},\Psi)
\end{equation}
where $\mathcal L_{\textrm{mat}}$ is the matter Lagrangian and $\Psi$ represents the matter fields. 

The EEP is verified with very high accuracy within the Solar System (see \cite{will:2006cq,will:2014la} and references therein). Amongst all the tests of the EEP, the search for spatial and temporal variations of the fundamental constants is a way to test the Local Position Invariance. Today, we have excellent constraints on the variations of the fine structure constant $\alpha$ \cite{damour:1996hc,fujii:2004ij,petrov:2006bs,webb:1999ao,webb:2001fv,webb:2003if,murphy:2001rt,murphy:2004hq,murphy:2003dz,king:2012hb,chand:2004fr,srianand:2004wo,webb:2011oj,tzanavaris:2005hb,tzanavaris:2007uk,curran:2011eq,marion:2003zr,bize:2003ly,rosenband:2008fk,peik:2004qf,ashby:2007dq,fortier:2007bh,fischer:2004ve,tobar:2010cr,guena:2012ys,olive:2004le,olive:2008fk}, on variations of the weak interaction constant $\alpha_W$ \cite{damour:1996hc,malaney:1993qq} and on the variation of the constants of strong interaction \cite{murphy:2004hq,fischer:2004ve,tobar:2010cr,guena:2012ys,dmitriev:2004dw,bedaque:2011ay} (for a review of all these tests, see \cite{uzan:2011vn}).

In GR, the gravitational interaction is mediated through one metric tensor only. Nevertheless, a lot of GR extensions consider the presence of additional fields (for a wide review of GR extensions, see \cite{clifton:2012fk}). In particular, following the work of Jordan, Brans and Dicke \cite{jordan:1949vn,brans:1961fk,brans:2014sc}, scalar-tensor theories of gravity have been widely studied in the literature. Originally, this type of theory has been studied with a minimal coupling between the matter fields and the scalar field. This means that there exists a metric such that the matter action can be written as (\ref{eq:matterAction}), while the scalar field modifies the dynamics of the metric\footnote{The representation in which the coupling (\ref{eq:matterAction}) appears is called the Jordan frame. Another representation widely used to study this kind of theory is the Einstein frame where the scalar and the tensor modes are kinematically decoupled. The Jordan and Einstein frame metrics are related by a conformal transformation \cite{damour:1992ys}.} \cite{damour:1992ys,damour:1990fk,damour:1993kx,damour:1993uq,damour:1993vn,hees:2012kx}. 

One general way to break the EEP is to introduce a nonminimal multiplicative coupling between the scalar field and matter fields, e.g.
\begin{equation}\label{eq:matterActionMod}
	S_{\textrm{mat}}=\sum_i \int d^4x \sqrt{-g} h_i(\phi)\mathcal L_i(g_{\mu\nu},\Psi_i)
\end{equation}
where the $h_i(\phi)$ are functions of the scalar field\footnote{All the functions $h_i$ can eventually be equal. Note that such universality allows the occurrence of an important cosmological convergence mechanism \cite{damour:1994fk}.} and $\mathcal L_i$ are the Lagrangians of the different matter fields. The dynamics of the scalar field and of the metric tensor are not important here and are encoded in the other part of the action named $S_{\textrm{grav}}(g_{\mu\nu},\phi)$.
 
Such a nonminimal coupling is motivated by several alternative theories such as the low energy action of string theories \cite{damour:1994fk,damour:1994uq,gasperini:2002kx,minazzoli:2014pb}, in the context of axions \cite{peccei:1977lr,dine:1981pb,kaplan:1985nb}, of generalized chameleons \cite{brax:2004fk,brax:2007xi,weltman:2008jb,ahlers:2008mq} \footnote{We use the term "generalized" to make the difference between the original chameleon papers \cite{khoury:2004fk} where the couplings are made through conformal transformations and not through a multiplicative coupling like the one considered here and in \cite{brax:2004fk}.}, by Kaluza-Klein theories with additional compactified dimensions \cite{fujii:2003fi,overduin:1997wb}, in the Bekentein-Sandvik-Barrow-Magueijo theory of varying $\alpha$ \cite{bekenstein:1982zr,sandvik:2002ly,barrow:2012wd,barrow:2013rr}\footnote{Although in this theory, the multiplicative coupling is not ``universal'' and is  restricted either to the kinetic part of the Lagrangian (see, e.g. \cite{sandvik:2002ly}), or to its interaction part (see, e.g. \cite{bekenstein:1982zr}), depending on the representation used (see \cite{hees:2014kq} for more details); but fermion masses are considered as independent of the scalar field \cite{bekenstein:1982zr,sandvik:2002ly,barrow:2012wd,barrow:2013rr}.} or in extended $f(R,\mathcal{L}_m)$ gravity \cite{harko:2013rz}. This type of coupling also appears in the context of the pressuron theory \cite{minazzoli:2013fk} characterized by $h_i\propto \sqrt{\phi}$  \footnote{In the pressuron theory, the scalar field naturally decouples in regions where the pressure is negligible \cite{minazzoli:2013fk,minazzoli:2014xz} and therefore naturally satisfies all Solar System tests of gravity.}.

It is  straightforward to show that this kind of coupling implies a variation of the fundamental constants. For example, since the fine-structure constant is related to the scalar field by $\alpha\propto h^{-1}_{\textrm{EM}}(\phi)$, its temporal variation is given by \cite{uzan:2011vn,damour:1994fk,damour:1994uq,damour:2002vn,damour:2002ys}
\begin{equation}\label{eq:alpha}
	\frac{\dot \alpha}{\alpha}=-\frac{h'_{\textrm{EM}}(\phi)}{h_{\textrm{EM}}(\phi)}\dot \phi
\end{equation}
where the dot corresponds to the temporal derivative, the prime to the derivative with respect to the scalar field $\phi$ and $h_{\textrm{EM}}$ is the coupling function appearing in front of the electromagnetic Lagrangian. 

In addition to variations of the fine structure constant, a coupling of the form (\ref{eq:matterActionMod}) in the electromagnetic sector implies a non-conservation of the photon number along geodesics \cite{minazzoli:2013zl}. Such a non-conservation can have several observational consequences. First of all, the expression of the luminosity distance is modified with respect to its GR expression \cite{minazzoli:2014xz}. Hence, one expects the distance-duality relation \cite{etherington:1933rm,etherington:2007sf,ellis:2007vn,ellis:2009fk,ellis:2013fk} to be modified accordingly. Therefore, there is a non-ambiguous relation between fine structure constant variations and violation of the distance-duality relation.

On the other hand, a non-conservation of the photon number should also modify the evolution of the CMB radiation \cite{lima:1996pr,lima:2000mn}. In particular, the cosmological evolution of the CMB temperature is affected by the coupling (\ref{eq:matterActionMod}). Therefore, there is also a link between variations of the fine structure constant and temperature-redshift relation violations. Moreover, the coupling (\ref{eq:matterActionMod}) also implies that the CMB radiation does not obey the adiabaticity condition \cite{lima:2000mn}, so that the CMB is not an equilibrium blackbody radiation. This situation is similar to what is obtained in tensor-scalar theory with disformal couplings \cite{van-de-bruck:2013uq,brax:2013dq}. As a consequence, the coupling (\ref{eq:matterActionMod}) produces a distortion of the CMB spectrum parametrized by a chemical potential $\mu$. This non-vanishing chemical potential can also be related to variation of the fine structure constant or to violation of the distance-duality relation.

In the framework of the action (\ref{eq:matterActionMod}), the four effects described previously (temporal variation of the fine structure constant, violation of the distance-duality relation, modification of the evolution of the CMB temperature and CMB spectral distortions) are closely related and are all linked to the evolution of $h_\textrm{EM}(\phi)$. In this paper, we will explore these links and show how they can be used to improve current constraints on some deviations from GR using constraints on other effects, and/or to explicitly test couplings of the form (\ref{eq:matterActionMod}), i.e. a wide range of different theories of gravity (including GR).

The paper is organised as follows. In section~\ref{sec:eta}, we derive the expression of the violation of the cosmic distance-duality from the action (\ref{eq:matterActionMod}) and we show that it can be expressed directly in terms of the coupling $h_\textrm{EM}(\phi)$. We also briefly review the experimental constraints on the violation of the distance-duality. In section~\ref{sec:dalpha}, we show how the temporal variation of the fine-structure constant is also related to the evolution of $h_\textrm{EM}(\phi)$ and we review the current experimental constraints on the variation of the fine structure constant. In section~\ref{sec:CMB}, we derive the evolution of the CMB temperature and the expression of the CMB chemical potential from first principles solving the Boltzmann equation for the distribution function. We also review the experimental constraints on the evolution of the CMB temperature and the limits on the chemical potential. In section~\ref{sec:exp}, we use the relations between the different observables in order to transform constraints on one type of observations into constraints on other types of observations. This is valid only for theories with a coupling (\ref{eq:matterActionMod}). We also use the different set of data simultaneously in order to test the coupling (\ref{eq:matterActionMod}). Indeed, any inconsistency between the data from two types of observations can be interpreted as a violation  of the coupling (\ref{eq:matterActionMod}). We show that currently no inconsistency is detected and we also discuss the improvements expected from the SKA or from EUCLID.

\section{Modification of the cosmic distance-duality relation}\label{sec:eta}
\subsection{Theoretical derivation}
The luminosity distance $D_L$ is operationally  defined by $D_L=\left(\frac{L}{4\pi F}\right)$ where $L$ is the luminosity of the source and $F$ is the observed flux of energy (see for example \cite{hobson:2006uq}). On the other hand the angular distance $D_A$ is defined by $D_A=\frac{\ell}{\Delta\theta}$ where $\ell$ is the proper size of the source and $\Delta\theta$ is the angular size of its observation \cite{hobson:2006uq}. In any space-time geometry and for any theory of gravity in which the reciprocity relation holds and the numbers of photons is conserved \cite{etherington:1933rm,etherington:2007sf,ellis:2007vn,ellis:2009fk,ellis:2013fk}, these two distances are related by the distance-duality relation 
\begin{equation}\label{eq:distDual}
 D_L(z)=(1+z)^2 D_A(z),
\end{equation}
where $z$ is the redshift of the source. The reciprocity relation is a purely geometric relation connecting area distances up and down the past light cone. This relation holds as long as photons propagate along null geodesics and the geodesic equation holds \cite{ellis:2009fk,ellis:2013fk}. Then, the assumption that the number of photons is conserved leads to the distance-duality relation (\ref{eq:distDual})\footnote{This term was first introduced in \cite{bassett:2004wm} to point the difference with reciprocity in gravitation theories other than GR.}. Violation of the distance-duality relation are parametrized by\footnote{All the papers in the literature used the definition of the parameter $\eta$ given by (\ref{eq:eta}) except in \cite{uzan:2004lp,ellis:2013fk} where the inverse is used $\tilde \eta=D_A(1+z)^2/D_L$.}
\begin{equation}\label{eq:eta}
 \eta(z)=\frac{D_L(z)}{D_A(z)(1+z)^2}.
\end{equation}

We will show that a coupling between a scalar field and the electromagnetic Lagrangian of the type (\ref{eq:matterActionMod}) modifies the distance-duality relation (\ref{eq:distDual}). Introducing the electromagnetic Lagrangian into the action (\ref{eq:matterActionMod}) and varying it with respect to the 4-potential $A^\mu$ leads to modified Maxwell equations\footnote{From now on, we will note $h(\phi)$ the coupling function related to electromagnetism $h_{\textrm{EM}}(\phi)$.}
\begin{equation}
 \nabla_\nu \left(h(\phi)F^{\mu\nu}\right)=0,
\end{equation} 
where $F^{\mu\nu}$ is the standard Faraday tensor.

The use of the geometric optics approximation consisting in expanding the 4-potential $A^\mu= \Re \left\{ \left(b^\mu + \epsilon c^\mu + O(\epsilon^2) \right) e^{i \theta / \epsilon} \right\}$ (see for example \cite{misner:1973fk}) leads to the usual null geodesic equation and to a modified conservation equation for the number of photons (the derivation of these expression can be found in \cite{minazzoli:2013zl,minazzoli:2014xz})
\begin{subequations}\label{eq:optic}
 \begin{eqnarray}
 k^\mu \nabla_\mu k^\alpha &=& 0\label{eq:nullGeo}\\
  k^\mu k_\mu&=&0\\
  \nabla_\nu\left(b^2 k^\nu\right)&=&-b^2k^\nu \partial_\nu \ln h(\phi)\label{eq:consN1}
 \end{eqnarray}
\end{subequations}
where  $k_\mu=\nabla_\mu\theta$ is the wave vector and $b$ the norm of $b^\mu$.  The fact that photons propagate on null geodesic means that the reciprocity relation holds \cite{ellis:2009fk,ellis:2013fk} but the violation of the conservation of the number of photons implies a violation of the distance-duality relation.

The integration of Eqs.~(\ref{eq:optic}) in a flat Friedmann-Lema\^itre-Robertson-Walker (FLRW) space-time leads to the expression of the luminosity distance (see \cite{minazzoli:2014xz} for a detailed derivation)
\begin{equation}
 D_L(z)=c(1+z) \sqrt{\frac{h(\phi_0)}{h(\phi(z))}} \int_0^z \frac{dz'}{H(z')}
\end{equation}
where $z$ is the cosmological redshift $1+z=\frac{a_0}{a}$ with $a$ the cosmic scale factor and the subscript $0$ stands for the present epoch ($\phi_0=\phi(z=0)$). On the other hand, the angular distance is a purely geometric feature that can be computed from the geodesic equation (see \cite{hobson:2006uq} for example). Therefore its expression is the same as in GR and is given by
\begin{equation}
 D_A(z)=\frac{c}{1+z}\int_0^z \frac{dz'}{H(z')}.
\end{equation}
The $\eta$ parameter characterizing scalar-tensor theories with a coupling (\ref{eq:matterActionMod}) is therefore given by
\begin{equation}\label{eq:etaM}
 \eta(z)=\sqrt{\frac{h(\phi_0)}{h(\phi(z))}}.
\end{equation}
Hence, the constraints on $\eta(z)$ can directly be interpreted as a constraint on the cosmological evolution of the scalar field.

\subsection{Experimental constraints}
Different kinds of observations have been used in order to constrain $\eta(z)$: Supernovae Ia data and observations of radio galaxies \cite{bassett:2004wm}, observations of clusters of galaxies \cite{holanda:2012yg,cao:2011uq,li:2013ov,uzan:2004lp,de-bernardis:2006kx,holanda:2010lq,holanda:2011fp,holanda:2012eu,goncalves:2012rw,li:2011bu,nair:2011ez,yang:2013gd}, Baryon Acoustic Oscillations and the Cosmic Microwave Background (CMB) \cite{lazkoz:2008ta,cardone:2012km},  the CMB spectrum \cite{ellis:2013fk} or  gamma-ray bursts \cite{holanda:2014rw}.	

Different parametrizations of $\eta(z)$ have been used in the literature in order to analyze cosmological observations. The most widespread ones are
\begin{subequations}\label{eq:eta_param}
	\begin{eqnarray}
		\eta(z)&=&\eta_0 \\
		\eta(z)&=&1 + \eta_1 z \\
		\eta(z)&=&1+\eta_2 ~\frac{z}{1+z} \\
		\eta(z)&=& 1+\eta_3~ \ln (1+z) \\ 
		\eta(z)&=& (1+z)^\varepsilon  \label{eq:parEps}.
	\end{eqnarray}
\end{subequations}

In Table~\ref{tab:estdotalpha}, we present the latest observational constraints on the parameters parametrizing $\eta(z)$. 

\begin{table}[htb]
\caption{Observational estimations of the parameters entering the expressions of $\eta(z)$ (\ref{eq:eta_param}) for $0\lesssim z \lesssim 8$ (depending on the study)  and derived estimation of the temporal variation of the fine structure constant.}
\label{tab:estdotalpha} 
\centering
\begin{tabular}{c c c c}
\hline
Ref. & Parameter  & Estimation & Derived est.  \\
&&& of $\dot\alpha/\alpha ~ [\times 10^{-11} \textrm{yr}^{-1}]$\\
\hline
\cite{lazkoz:2008ta}   & $\eta_0$      &$0.95\pm 0.025$            & -\\
\cite{cao:2011uq}      & $\eta_0$      &$0.97_{-0.06}^{+0.05}$     & - \\
\cite{cao:2011uq}      & $\eta_1$      &$-0.01_{-0.16}^{+0.15}$    & $0.16\pm2.6$\\
\cite{cardone:2012km}  & $\eta_1$      & $-0.273\pm0.125$          & $4.4\pm2.01$\\
\cite{holanda:2012yg}  & $\eta_1$      & $-0.06\pm 0.08$           & $0.97\pm1.3$ \\
\cite{yang:2013gd}     & $\eta_1$      &$0.02_{-0.17}^{+0.2}$      & $-0.32\pm3.2$ \\
\cite{cao:2011uq}      & $\eta_2$      &$-0.01_{-0.24}^{+0.21}$    & $0.16\pm3.9$  \\
\cite{holanda:2012yg}  & $\eta_2$      & $-0.07\pm 0.12$           & $1.1\pm1.9$\\
\cite{cao:2011uq}        & $\eta_3$      & $-0.01_{-0.19}^{+0.22}$           & $0.16\pm 3.5$\\
\cite{li:2013ov}       & $\varepsilon$ & $0.066^{+0.037}_{-0.035}$ & $-1.1\pm0.6$ \\
\cite{avgoustidis2009jc} & $\varepsilon$ & $-0.01_{-0.09}^{+0.08}$        &  $0.16 \pm 1.42$ \\
\cite{avgoustidis:2010jc} & $\varepsilon$ & $-0.04_{-0.07}^{+0.08}$        &  $0.63 \pm 1.26$ \\
\cite{holanda:2014rw}  & $\varepsilon$ & $0.020 \pm  0.055$        & $-0.32\pm0.89$\\
\hline
\end{tabular}
\end{table}

\section{Temporal variation of the fine structure constant}\label{sec:dalpha}
\subsection{Theoretical derivation}
Since $\alpha\propto h^{-1}(\phi)$ \cite{damour:1994fk,damour:2002vn}, the temporal variation of $\alpha$  can  be related to the function $\eta(z)$. More precisely, one has
\begin{equation}\label{eq:dalpha_eta}
	\frac{\Delta \alpha(z)}{\alpha}=\frac{\alpha(z)-\alpha_0}{\alpha_0}=\frac{h(\phi_0)}{h(\phi)}-1=\eta^2(z)-1.
\end{equation}

This shows that for the class of theory considered in this paper, a violation of the distance-duality relation is directly linked to a violation of the EEP. In particular, experimental constraints on the function $\eta(z)$ can be transposed into a constraint on the temporal variation of $\alpha$ and inversely.

Taking the derivative at the current epoch of (\ref{eq:dalpha_eta}) leads to
\begin{equation}
 \left.\frac{\dot\alpha}{\alpha}\right|_0=-2 H_0 \left.\frac{d\eta}{dz}\right|_0
\end{equation}
where $H_0$ is the Hubble constant at the present time. If one uses the parametrizations of $\eta(z)$ from (\ref{eq:eta_param}), the last expression becomes
\begin{equation}
 -\frac{1}{2H_0}\left.\frac{\dot\alpha}{\alpha}\right|_0 = \eta_1=\eta_2=\eta_3=\varepsilon.	\label{eq:relatalphaeta1}
\end{equation}

\subsection{Experimental constraints}\label{sec:dalphaExp}
Currently, the best laboratory constraint on the time variation of the fine-structure constant is given by \cite{rosenband:2008fk}
\begin{equation}\label{eq:constdotalpha}
\left.\frac{\dot\alpha}{\alpha}\right|_0=(-1.6\pm2.3)\times 10^{-17}	\textrm{yr}^{-1}.
\end{equation}

Now, variations of $\alpha$ over a longer time can also be considered. Bounds on $\Delta \alpha/\alpha$ can be derived from the CMB data at $z\approx 10^3$ \cite{planck-collaboration:2013fk} and from Big Bang Nucleosynthesis at $z\approx 10^9$ \cite{iocco:2009la} but are not very stringent (see Table~\ref{tab:dalpha}). Observational searches for varying $\alpha$ have also used absorption systems in the spectra of distant quasars \cite{webb:1999ao, srianand:2004wo}. Evidence of a variation of $\alpha$ has been found using the Keck telescope \cite{webb:1999ao}
for $z$ between 0.2 and 4.2. A null result has been obtained considering observations from the Very Large Telescope \cite{srianand:2004wo} but this conclusion might suffer from biases in the data analysis \cite{murphy:2007mz,murphy:2008pd}. A review of the constraints on $\Delta \alpha/\alpha$ can be found in \cite{uzan:2011vn,chiba:2011gd}. Some of the important results are summarized in table~\ref{tab:dalpha}.
More recently, further evidence of a deviation of $\alpha$ from its current value has been found \cite{webb:2011oj,king:2012hb} using Keck and VLT observations. Nevertheless, two different values have been found for the two data sets: for the VLT, it is found that $\Delta\alpha/\alpha=(0.208\pm0.124)\times 10^{-5}$ \cite{king:2012hb} while for the Keck observations, it is found that $\Delta \alpha/\alpha=(-0.6\pm0.22)\times10^{-5}$ \cite{king:2012hb}. The results seem to depend on which hemisphere is considered, suggesting a dipolar dependence of $\alpha$ in the sky \cite{webb:2011oj,king:2012hb}. 

\begin{table}[htb]
\caption{Observational constraints on the temporal variations of the fine structure constant.}
\label{tab:dalpha} 
\centering
\begin{tabular}{c c c c}
\hline
 Observation & Ref.   & z & Estimation   \\\hline
 Oklo reactor & \cite{gould:2006zl} & 0.16 & $(6.5\pm 8.7)\times 10^{-8}$ \\
 Quasar abs. lines & \cite{webb:2001fv} & 0.5-3.5 & $(-0.72 \pm 0.18)\times 10^{-5}$\\
 Quasar abs. lines (VLT) & \cite{srianand:2004wo} & 0.4-2.3 & $(-6\pm 6)\times 10^{-7}$\\
 Quasar abs. lines (Keck) & \cite{murphy:2004hq} & 0.2-4.2 & $(-5.7\pm 1.1)\times 10^{-6}$ \\
 Quasar abs. lines (VLT) & \cite{king:2012hb} & 0.2-3.6 & $(2.08\pm 1.24)\times 10^{-6}$ \\
 CMB & \cite{planck-collaboration:2013fk} & $10^3$ & $(8\pm 20)\times 10^{-3} $ \\
 BBN & \cite{avelino:2001ty} & $10^{10}$ & $(-7\pm 5)\times 10^{-3}$\\
\hline
\end{tabular}
\end{table}

\section{CMB temperature and distortions}\label{sec:CMB}
\subsection{Theoretical derivation}
In this section, we will derive the evolution of the temperature of the CMB using an approach based on the kinetic theory (see chapter 4 of \cite{peter:2009kl} and chapter 4 of \cite{durrer:2008cr}).

First of all, let us notice that Eq.~(\ref{eq:consN1}) in a flat FLRW space-time can be written in terms of the number of photons $n\propto k^0 b^2$ \cite{misner:1973fk}
\begin{equation}\label{eq:consN}
 \dot n+ 3Hn=-n \frac{\partial \ln h(\phi(t))}{\partial t},
\end{equation}
where $t$ is the proper time along matter worldlines.
This equation gives the evolution of the number of photons along a single light ray.

From a microscopic perspective, we define the distribution function $f$ of a fluid of photons. The evolution of this distribution function satisfies a Boltzmann equation (see Section 4.1. of \cite{durrer:2008cr})
\begin{equation}
  \mathcal L f = \frac{df}{d\lambda}=  \tilde p^\alpha \frac{\partial f}{\partial x^\alpha} +  \frac{d\tilde p^i}{d\lambda} \frac{\partial f}{\partial \tilde p^i}=\mathcal C[f]
\end{equation}
with $\mathcal L$ the Liouville operator, $\tilde p^\mu$ the coordinates of the 4-impulsion in the coordinate basis and  $\mathcal C[f]$  an effective collision term present because of the coupling between the scalar field and the electromagnetic Lagrangian. Since we have shown that at the eikonal approximation, photons still follow null geodesics (\ref{eq:nullGeo}), we have
\begin{equation}
   \mathcal L f = \frac{df}{d\lambda}=  \tilde p^\alpha \frac{\partial f}{\partial x^\alpha} -\Gamma^i_{\mu\nu}\tilde p^\mu \tilde p^\nu \frac{\partial f}{\partial \tilde p^i}=\mathcal C[f].
\end{equation}
In the case of the FLRW geometry, the last equation is standard (see for example section 4.1. of \cite{durrer:2008cr}) and becomes
\begin{equation}\label{eq:liouville}
  \mathcal L f = \frac{df}{d\lambda}=p^0\frac{\partial f}{\partial t}-H p^0 p \frac{\partial f}{\partial p}=\mathcal C[f]
\end{equation}
where $p^\mu$ are the coordinates of the 4-impulsion in a local tetrad. In particular, since $p^\mu p_\mu=0$ for photons, in a local tetrad we have $p^0=p$ where $p$ is the standard Euclidean norm of the 3-vector $(p^1,p^2,p^3)$.

The integration of (\ref{eq:liouville}) in the case of a directional radiation characterized by $f\propto \delta^{(3)}(\vec p-\vec p_0)$ should lead to the equation of non-conservation of the number of photons in a electromagnetic radiation (\ref{eq:consN}). This allows one to identify the collision term which is given by
\begin{equation}
 \mathcal C[f]=-p f \partial_t \ln h(\phi).
\end{equation}
The Liouville equation becomes
\begin{equation}\label{eq:liouville2}
 \mathcal L f = \frac{df}{d\lambda}=p\frac{\partial f}{\partial t}-H p^2  \frac{\partial f}{\partial p}=-p f \partial_t \ln h(\phi).
\end{equation}
In a homogeneous and spherically symmetric case, the distribution function depends only on $t$ and $p$: $f(t,p)$.

The number of massless particles and their mean energy density are defined from a microscopic perspective as \cite{durrer:2008cr}
\begin{subequations}\label{eq:defVar}
 \begin{eqnarray}
  n&=&\frac{N_B}{(2\pi)^3}\int f(t,p) d^3p \\
  \rho &=&\frac{N_B}{(2\pi)^3}\int p f(t,p) d^3p. \label{eq:defrho}
 \end{eqnarray}
\end{subequations}
with $N_B$ the degeneracy factor for the particles which is 2 in the case of photons.
Therefore, the integration of the Liouville equation (\ref{eq:liouville2}) leads to equations of conservation of the number of photons and of the energy density of the photons
\begin{subequations}\label{eq:cons}
 \begin{eqnarray}
  \dot n+3Hn&=& -n \partial_t \ln h(\phi) =\Psi \\
  \dot \rho+4H\rho&=& -\rho \partial_t \ln h(\phi) = C_x
 \end{eqnarray}
\end{subequations}
where the terms $\Psi$ and $C_x$ are introduced to compare our results with \cite{lima:1996pr,lima:2000mn}. As one can see, any theory with a coupling like the one considered in the action (\ref{eq:matterActionMod}) does not satisfy the adiabaticity condition (given by Eq.~(11) of \cite{lima:2000mn})
\begin{equation}\label{eq:adiab}
 C_x=\frac{\rho}{n}\Psi \neq\frac{4\rho}{3n}\Psi.
\end{equation}
Therefore, making the assumption of adiabaticity as, for example, in \cite{avgoustidis:2013gd} for couplings of the form (\ref{eq:matterActionMod}) is not justified. The coupling (\ref{eq:matterActionMod}) implies that the CMB radiation is not an equilibrium blackbody radiation. This is similar to what appears in tensor-scalar theory with disformal coupling \cite{van-de-bruck:2013uq,brax:2013dq}. This is due to the fact that the distribution function $f$ is not conserved (\ref{eq:liouville2}). A way to parametrize the deviation from the blackbody spectrum is to introduce a chemical potential $\mu$ (see Section 8.2 of \cite{durrer:2008cr}). The distribution function can be written as
\begin{equation}
 f(t,p)=\frac{1}{e^{p/T+\mu}-1}
\end{equation}
where the temperature and the chemical potential depend on the cosmological evolution. If we introduce this expression in the definition of the number of particles $n$ and the energy density $\rho$ (\ref{eq:defVar}), we get, at first order in $\mu$ (we will see that experimental limits on $\mu$ impose $\mu< 10^{-4}$)
\begin{subequations}\label{eq:nrho}
 \begin{eqnarray}
  n&=&\frac{2\zeta(3)T^3}{\pi^2}\left(1-\frac{\pi^2}{6\zeta(3)}\mu\right) \\
  \rho&=&\frac{\pi^2T^4}{15}\left(1-\frac{90\zeta(3)}{\pi^4}\mu\right)\label{eq:rhoT}
 \end{eqnarray}
\end{subequations}
where $\zeta(x)$ is the Riemann zeta-function. Since the deviations from the GR case induced by the coupling $h(\phi)$ are expected to be small, we can use an expansion of the quantities
\begin{subequations}
 \begin{eqnarray}
  T&=&T_{(0)}+\delta T\\
  n&=&n_{(0)}+\delta n \\
  \rho &=& \rho_{(0)} +\delta \rho 
 \end{eqnarray}
\end{subequations}
where $x_{(0)}$ refers to the value of $x$ in GR (when $h(\phi)=1$) while $\mu$ is already a first-order term. Introducing this expansion and solving the Eqs.~(\ref{eq:nrho}) at first order leads to (for a detailed derivation, see section 8.2 of \cite{durrer:2008cr})
\begin{subequations}\label{eq:pert}
 \begin{eqnarray}
 \T0 &=&\left(\frac{15 \rhou}{\pi^2}\right)^{1/4}=\left(\frac{\pi^2 \n0}{2\zeta(3)}\right)^{1/3}\\
  \frac{\delta T}{T_{(0)}}&=&\frac{\frac{\delta\rho}{\rhou} - \frac{540\zeta(3)^2}{\pi^6} \frac{\delta n}{\n0} }{4\left(1-\frac{405\zeta(3)^2}{\pi^6}\right)} \\
  \mu&=&\frac{3\zeta(3)}{2\pi^2}\frac{3\frac{\delta \rho}{\rhou}-4\frac{\delta n}{ {\n0}}}{1-\frac{405\zeta(3)^2}{\pi^6}}.
 \end{eqnarray}
\end{subequations}
In particular, we can see that $\mu=0$ is obtained in the case where the adiabaticity condition (\ref{eq:adiab}) is satisfied.
The exact solutions of Eqs.~(\ref{eq:cons}) are given by 
\begin{eqnarray}
na^3 h(\phi) &=& n_ia^3_i  h(\phi_i), \\
\rho a^4 h(\phi) &=& \rho_ia^4_i  h(\phi_i)
\end{eqnarray}
 where the indices $i$ refer to some initial conditions. At zeroth order, this gives the usual GR behaviour $n_{(0)}\propto a^{-3}$ and $\rhou\propto a^{-4}$. At first order, we have
\begin{equation}
 \frac{\delta n}{\n0}=\frac{\delta\rho}{\rhou}=\frac{h(\phi_\textrm{CMB})}{h(\phi)}-1\equiv \delta h(\phi).
\end{equation}
The choice of initial conditions at $t_i=t_{\textrm{CMB}}$ concording with the CMB is required if we want the chemical potential to vanish at that time. This is consistent with assuming that the CMB radiation is initially emitted as a blackbody. Using (\ref{eq:dalpha_eta}), we can express $\delta h$ as
\begin{equation}\label{eq:deltah}
 \delta h(\phi)=\frac{\eta^2(z)}{\eta^2(z_\textrm{CMB})}-1=\frac{\Delta\alpha(z)}{\alpha}-\frac{\Delta \alpha(z_\textrm{CMB})}{\alpha}.
\end{equation}

Inserting this result in Eqs.~(\ref{eq:pert}) gives
\begin{subequations}
 \begin{eqnarray}
 \T0&=&\frac{T_ia_i}{a}=T_i\frac{1+z}{1+z_i} \\
  \frac{\delta T}{\T0}&=&\frac{1 - 540\zeta(3)^2/\pi^6  }{1-405\zeta(3)^2/\pi^6}\frac{\delta h(\phi)}{4} \approx 0.1204\, \delta h(\phi)\\
  \mu &=&\frac{3\zeta(3)}{2\pi^2}\frac{\delta h(\phi)}{\frac{405\zeta(3)^2}{\pi^6}-1}\approx -0.4669 \, \delta h(\phi).
 \end{eqnarray}
\end{subequations}
Therefore, the temperature is given by
\begin{eqnarray*}
 T&=&T_i \frac{1+z}{1+z_i}\left( 1+0.12 \delta h(\phi(z)) \right)\\
 &=&T_0(1+z) \Big[1+ 0.12\big(\delta h(\phi(z))-\delta h(\phi(0) \big) \Big]
\end{eqnarray*}
where the subscript 0 stands for values at $z=0$. Using (\ref{eq:deltah}) and keeping the leading term in $\Delta \alpha/\alpha$ and in $\eta^2-1$, one gets
\begin{subequations}
\begin{eqnarray}
 T(z)&=&T_0(1+z)\Big[1+0.12\frac{\Delta \alpha(z)}{\alpha}\Big]\label{eq:Talpha}\\
 &=&T_0(1+z)\Big[0.88+0.12\eta^2(z)\Big].
\end{eqnarray}
\end{subequations}
This relation makes a very precise link between a deviation of the cosmic evolution of the CMB temperature, a temporal evolution of the fine structure constant and a violation of the cosmic distance-duality. This relation is different from the one obtained in \cite{barrow:2013rr}. The reason comes from the fact that in \cite{barrow:2013rr}, the density is supposed to be related to the temperature as $\rho \propto T^4$. This means that they implicitly suppose that $\mu=0$ as can be seen from  Eq.~(\ref{eq:rhoT}) or equivalently that the CMB radiation still follows a blackbody spectrum which is not the case since the adiabaticity condition is violated (\ref{eq:adiab}).

Similarly, we have a direct concordance between the CMB spectral distortions parametrized by $\mu$, the variation of the fine structure constant and the corresponding violation of the distance-duality relation
\begin{eqnarray}\label{eq:mu}
 \mu&=&0.47\left(1-\frac{1}{ \eta^2(z_\textrm{CMB})}\right) =0.47\frac{\Delta \alpha(z_\textrm{CMB})}{\alpha}\\
&=&3.92\left(\frac{T(z_\textrm{CMB})}{T_0(1+z_\textrm{CMB})}-1\right).\nonumber
\end{eqnarray}
This expression differs from the one found in \cite{brax:2013dq}. The reason comes from the fact that in \cite{brax:2013dq}, the temperature is supposed to follow the standard evolution $T\propto (1+z)$ in the calculation of $\mu$.

\subsection{Experimental constraints}
First of all, a constraint on the CMB distortions has been obtained by COBE/FIRAS \cite{fixsen:1996dp}
\begin{equation}\label{eq:constMu}
 \left|\mu\right|<9\times 10^{-5}
\end{equation}
at 95 \% of confidence.

Usually, the experimental constraints on the evolution of the temperature are expressed in function of the parameter $\beta$ defined by
\begin{equation}\label{eq:betaTemp}
 T(z)=T_0(1+z)^{1-\beta}.
\end{equation}
Observations of the Sunyaev-Zel'dovich effect and measurements of molecular species absorptions have led to estimations of $\beta$ given in table~\ref{tab:beta}.

\begin{table}[htb]
\caption{Observational estimations of $\beta$ which parametrizes the evolution of the CMB temperature (\ref{eq:betaTemp})}
\label{tab:beta} 
\centering
\begin{tabular}{c c }
\hline
Ref. &   Estimation of $\beta$  \\
\hline
\cite{hurier:2014aa}   & $ 0.006 \pm 0.013$    \\
\cite{saro:2014mn}      & $0.005 \pm 0.012$    \\
\hline
\end{tabular}
\end{table}

\section{Experimental constraints}\label{sec:exp}
In the previous section, we have shown that four important cosmological observables are directly related to each other in the framework of the coupling (\ref{eq:matterActionMod}). Basically, temporal variations of the fine-structure constant, violation of the cosmic distance-duality relation and the evolution of the CMB temperature are all related to the evolution of the function $h(\phi)$ through:
\begin{equation}\label{eq:hphi}
 \frac{h(\phi_0)}{h(\phi(z))}=\eta^2(z)=\frac{\Delta\alpha(z)}{\alpha}+1=8.33\frac{T(z)}{T_0(1+z)}-7.33.
\end{equation}
Furthermore, the actual chemical potential $\mu$ of the CMB spectrum is also related to the previous quantities at $z=z_\textrm{CMB}$ by the relation (\ref{eq:mu}).

There are two different ways of using these relations. First, if we assume that the coupling between the scalar field and the EM Lagrangian can be written as (\ref{eq:matterActionMod}), we can use the relations between the different observables to constrain some of them by using measurements of other types of observations. As we will see below, the constraints on the variations of the fine structure constant are the most competitive. Therefore, we can transform them to obtain improved constraints on violations of the distance-duality and on the evolution of the temperature. We stress out that this procedure applies only if the coupling (\ref{eq:matterActionMod}) is correct. In particular, this is thus valid for GR (where $h(\phi)=1$), but also for all theories conformally coupled to a metric.

On the other hand, the observations from different data sets can be combined together in order to search for a hypothetical violation of the coupling (\ref{eq:matterActionMod}). Indeed, a violation of the relations (\ref{eq:hphi}) observed with two different data sets would imply that the coupling (\ref{eq:matterActionMod}) is not the one that describes Nature. One such evidence would be particularly important since it would rule out all the theories of gravity where this coupling appeared, including GR and many others.

\subsection{Transformations of the experimental constraints assuming a multiplicative coupling holds}
In this section, we will assume that the coupling (\ref{eq:matterActionMod}) holds and we will use the different relations between the observables in order to improve the constraints on some of them.

\subsubsection{Transformation between $\eta$ and $\Delta\alpha/\alpha$}
First of all, the relation between $\eta$ and $\Delta\alpha/\alpha$ (\ref{eq:dalpha_eta}) allows one to transform constraints on $\eta$ into constraints on variations of the fine structure constant and inversely. We transform the experimental constraints on $\eta$ into a constraint on the current variation of the fine structure constant $\left.\dot\alpha/\alpha\right|_0$. For this, we use the relation (\ref{eq:relatalphaeta1}) and the estimation of the Hubble constant provided by the Planck data $H_0=78.8\pm0.77 \ \textrm{km/s/MPc}$ \cite{planck-collaboration:2013fk}. The resulting estimations of $\dot\alpha/\alpha$ are given in table~\ref{tab:estdotalpha}. One can see that the obtained constraints are 6 orders of magnitude larger than the current constraint on $\dot\alpha/\alpha$ obtained by a laboratory experiment (\ref{eq:constdotalpha}). 

On the other hand, we can use the laboratory constraint (\ref{eq:constdotalpha}) in order to estimate the parameters entering the standard parametrization of $\eta(z)$. This leads to
\begin{equation}\label{eq:constEtaAlpha}
\eta_1=\eta_2=\eta_3=\varepsilon=(10 \pm14)\times10^{-8}. 
\end{equation}
This means that the current null result on a temporal variation of the fine structure constant locally constrains the present derivative of $\eta(z)$ by 6 orders of magnitude better than using cosmological observations.  Note that the constraint (\ref{eq:constEtaAlpha}) also applies for parametrizations where $\eta_0$ is not forced to be 1 (ie. for $\eta(z)=\eta_0+ \eta_i f_i(z)$, instead of $\eta(z)=1+ \eta_i f_i(z)$ as in (\ref{eq:eta_param})).

\begin{figure*}[hbt]
\begin{center}
\includegraphics[width=0.95\textwidth]{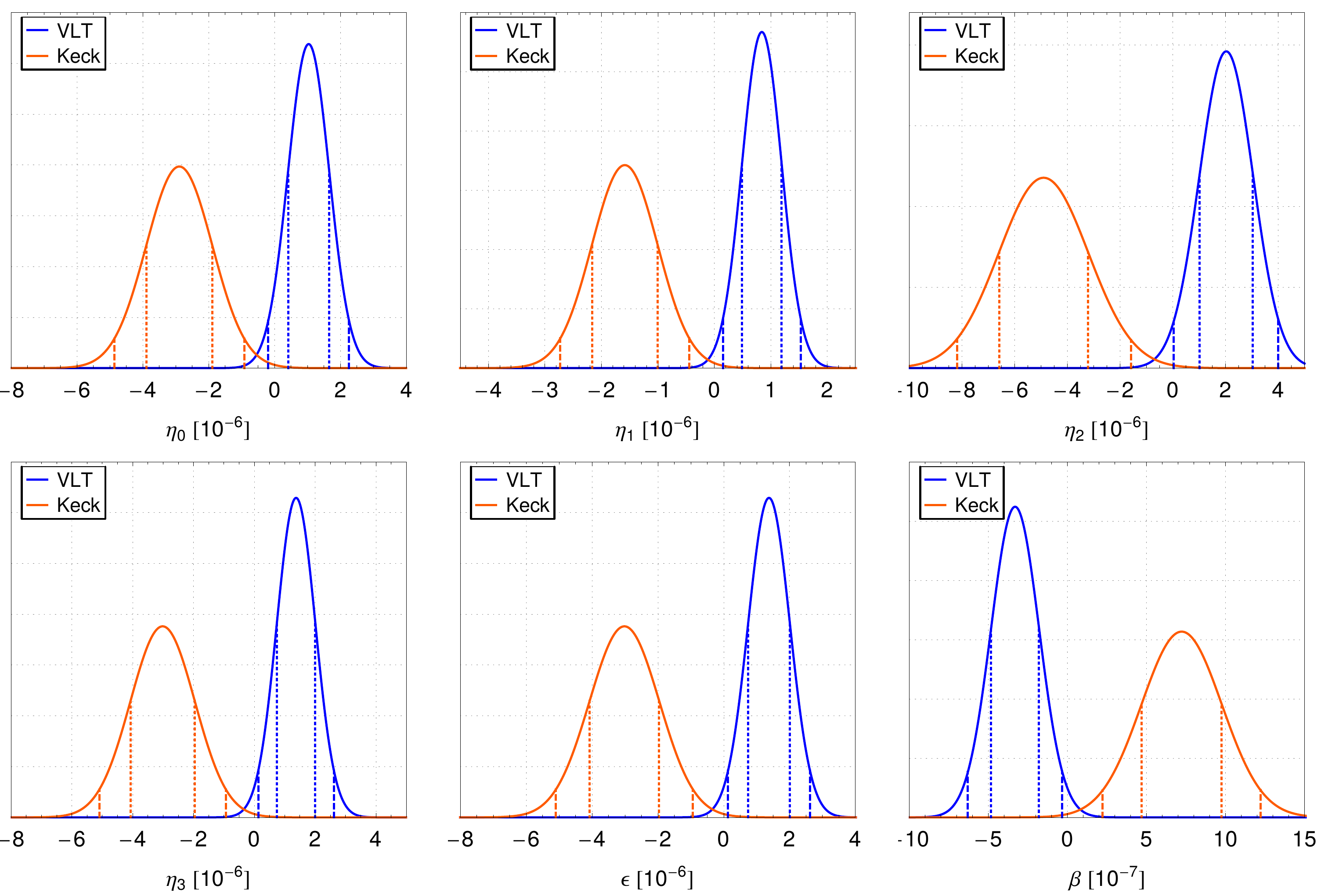}
\end{center}
\caption{Posterior probability densities of the parameters $\eta_i$, $\varepsilon$ that are parametrizing $\eta(z)$ (\ref{eq:eta_param}) and $\beta$ that is parametrizing the evolution of the temperature (\ref{eq:betaTemp}). The Bayesian inversion is done from $\Delta\alpha/\alpha$ data coming from VLT (Southern hemisphere) and Keck observatory (Northern hemisphere) assuming the relations (\ref{eq:dalpha_eta}) and (\ref{eq:Talpha}) hold. The dotted lines represent the 68 \% confidence intervals while the dashed lines represent the 95 \% confidence intervals.}
\label{fig:Post}
\end{figure*}

The constraint (\ref{eq:constEtaAlpha}) is very impressive but it relies on only one observation at $z=0$. Therefore, it is also interesting to apply the same procedure using constraints on variations of the fine structure constant at different redshifts. We used observations of absorption lines of  quasars to estimate the parameters characterizing $\eta(z)$ (\ref{eq:eta_param}). We used values of $\Delta\alpha/\alpha$ from 154 absorbers observed with the VLT (this data set can be found in \cite{king:2012hb}) and values from 128 absorbers observed at the Keck observatory (this data set can be found in \cite{murphy:2003dz}). We have performed a Bayesian estimation of the parameters $\eta_i$ and $\varepsilon$ that are parametrizing the $\eta(z)$ function (\ref{eq:eta_param}) using the relation (\ref{eq:dalpha_eta}). The posterior probability densities are presented on figure~\ref{fig:Post}. Table~\ref{tab:esteta} lists the corresponding estimations using the different data sets. Evidence of deviations of the parameters from their GR values are found with the two data sets separately. Nevertheless, the two data sets are incompatible. This is due to the fact that the variation of the fine structure constant is different in the Northern and in the Southern hemispheres \cite{webb:2011oj,king:2012hb} (see also the discussion in section~\ref{sec:dalphaExp}). Therefore, if the electromagnetic Lagrangian is coupled to a scalar field by a coupling of the type (\ref{eq:matterActionMod}), $\Delta\alpha/\alpha$ observations predict a violation of the distance-duality relation  at the $10^{-6}$ level, which should, however, be different in both hemisphere (independently of the coupling function).

\begin{table}[htb]
\caption{Values of the parameters entering the expression of $\eta(z)$ (\ref{eq:eta_param}) estimated using $\Delta \alpha/\alpha$ data from VLT \cite{king:2012hb} and from the Keck Observatory \cite{murphy:2003dz} assuming relation (\ref{eq:dalpha_eta}) holds.}
\label{tab:esteta} 
\centering
\begin{tabular}{c c c }
\hline
 Parameter  & \multicolumn{2}{c}{Estimation $[\times 10^{-7}]$} \\
 & VLT & Keck  \\
\hline
$\eta_0-1$     &  $10\pm 6\phantom{0.}$ &  $-29\pm10$             \\
$\eta_1$       &  $8.4\pm 3.5$           &  $-16\pm6\phantom{1}$   \\
$\eta_2$       &  $20\pm 10\phantom{.}$  &  $-49\pm17$             \\
$\eta_3$       &  $14\pm 6\phantom{0.}$  &  $-30\pm11$             \\
$\varepsilon$  & $14\pm 6\phantom{0.}$   &  $-30\pm 11$            \\
\hline
\end{tabular}
\end{table}

\subsubsection{Transformation between $\Delta\alpha/\alpha$ and the CMB temperature}
As shown in the previous section, the constraints on the variations of the fine structure constant are far more stringent than the ones on the violations of the distance-duality relation. We can thus use the constraints on the current temporal variation of the fine structure constant to constraint $\beta$ which parametrizes the evolution of the CMB temperature (\ref{eq:betaTemp}). For this, we need to derive the relation (\ref{eq:Talpha}) 
\begin{equation}
 \left.\frac{dT}{dz}\right|_{z=0}=T_0\left(1-0.12 H_0 \left.\frac{\dot\alpha}{\alpha}\right|_0\right)
\end{equation}
which becomes, after introducing the parametrization (\ref{eq:betaTemp})
\begin{equation}
 \beta=\frac{0.12}{ H_0} \left.\frac{\dot\alpha}{\alpha}\right|_0 .
\end{equation}
Using this equation, the value and uncertainty on $H_0$ from the Planck data \cite{planck-collaboration:2013fk} and the laboratory constraint on $\dot\alpha/\alpha$ (\ref{eq:constdotalpha}), we get
\begin{equation}
 \beta=(-2.4\pm 3.4)\times 10^{-8}.
\end{equation}
This constraint improves the one coming from current direct observations of the CMB temperature (see table~\ref{tab:beta}) by 7 orders of magnitude. 

The last constraint is very impressive but, once again, it relies on only one observation at $z=0$. One can also look at constraints on variations of the fine structure constant at different redshift $z$. We use the same VLT and Keck data as in the previous section in order to do a Bayesian estimation of $\beta$ from $\Delta \alpha/\alpha$ data by using the relation
\begin{equation}\label{eq:beta_dalpha}
 (1+z)^{-\beta}=1+0.12 \frac{\Delta\alpha(z)}{\alpha}.
\end{equation}
The posterior probability density is presented on figure~\ref{fig:Post} (bottom right) and the corresponding estimations are given in table~\ref{tab:estbeta}. Once again, the obtained estimations present a deviation from the GR values at the level of $10^{-6}$ but the two data sets are not compatible. This is due to the fact that the variation of the fine structure constant is different in both hemispheres \cite{webb:2011oj,king:2012hb}. Therefore, if the coupling (\ref{eq:matterActionMod}) holds, the observations of temporal variations of the fine structure constant suggests a deviation of the evolution of the CMB temperature 5 orders of magnitude smaller than current direct observation of the CMB temperature capabilities.

\begin{table}[htb]
\caption{Values of the parameters entering the expression of $T(z)$ (\ref{eq:betaTemp}) estimated using $\Delta \alpha/\alpha$ data from VLT \cite{king:2012hb} and from the Keck Observatory \cite{murphy:2003dz} assuming the relation (\ref{eq:beta_dalpha}) holds.}
\label{tab:estbeta} 
\centering
\begin{tabular}{c c c }
\hline
 Parameter  & \multicolumn{2}{c}{Estimation $[\times 10^{-7}]$} \\
 & VLT & Keck  \\
\hline
$\beta$     &  $-3.3\pm 1.5$           &  $7.2\pm 2.5$             \\
\hline
\end{tabular}
\end{table}

\subsubsection{CMB distortions}
Finally, the relation (\ref{eq:mu}) allows one to transform the constraint on $\mu$ into a constraint on $\Delta\alpha(z_\textrm{CMB})/\alpha$. Using the constraint (\ref{eq:constMu}) and the relation (\ref{eq:mu}), we derive a constraint on the temporal variation of the fine structure constant
\begin{equation}
 \left|\frac{\Delta\alpha(z_\textrm{CMB})}{\alpha}\right|<1.91\times 10^{-4}.
\end{equation}
Let us remember that the constraint on $\Delta\alpha(z_\textrm{CMB})/\alpha$ coming from an analysis of the CMB anisotropies with Planck data is at the level of $10^{-3}$ only (see table~\ref{tab:dalpha}).

\subsection{Test of the multiplicative coupling}
In the previous section, we have shown how to use the relations between the variations of the fine structure constant, violation of the distance-duality relation, evolution of the CMB temperature and the CMB distortions in order to translate the measurements from one type of observations into the other types. As clearly stated, this can be done only if the coupling (\ref{eq:matterActionMod}) holds.

We can also use the different sets of data to assess the validity of the coupling (\ref{eq:matterActionMod}). Indeed, if the measurements coming from two different types of observations (e.g. between $\Delta\alpha/\alpha$ and $T_\textrm{CMB}$ or between $\Delta\alpha/\alpha$ and $\eta$) indicate a violation of their corresponding relation, this would be an indication of a violation of the coupling (\ref{eq:matterActionMod}). This kind of test is able to rule out couplings of the form (\ref{eq:matterActionMod}) and is therefore quite important since this kind of coupling generically appears in numerous alternative theories of gravity such as in perturbative string theory \cite{damour:1994fk,damour:1994uq,gasperini:2002kx}, Kaluza-Klein theories \cite{fujii:2003fi,overduin:1997wb}, axion theory \cite{peccei:1977lr,dine:1981pb,kaplan:1985nb}, BSBM theory \cite{bekenstein:1982zr,sandvik:2002ly,barrow:2012wd,barrow:2013rr}, \dots

In this work, to assess if the different observations are consistent with a coupling of the type (\ref{eq:matterActionMod}), we use the relations (\ref{eq:hphi}). Basically, we transform constraints on $\Delta\alpha/\alpha$, on $\eta(z)$ and on the CMB temperature into a constraint on $h(\phi)/h(\phi_0)$. We therefore suppose implicitly that the coupling (\ref{eq:matterActionMod}) holds. Then, we compare the different constraints on $h(\phi)/h(\phi_0)$ coming from different types of observations to see if they are consistent. Any inconsistency would be a signature of a deviation from the type of coupling (\ref{eq:matterActionMod}) independently of the coupling function $h(\phi)$.

We analyze the different data using Gaussian Processes (GP) with the software GaPP (Gaussian Processes in Python) \cite{seikel:2012fk}. GP provide a model-independent smoothing technique\footnote{In the sense that they do not introduce any uncontrolled physical assumptions. They do, however, suppose that data follow Gaussian distributions}. They are described in detail in \cite{seikel:2012fk} (see also \cite{seikel:2012uq,bilicki:2012kx,Bester:2013fya,yahya:2014ys} for other uses of GaPP in a cosmological context). Instead of assuming a particular form of the reconstructed function, GP consider typical changes of the function. They are parametrized by a covariance function which depends on two hyperparameters: $\ell$ which corresponds to a typical distance one needs to move in the input space to observe a significant change in the function and $\sigma_f^2$ which is a typical change of the function. In this paper, the covariance function used is the standard squared exponential function (other covariance functions have been tried and they do not significantly alter the results).

\begin{figure*}[hbt]
\begin{center}
\includegraphics[width=0.48\textwidth]{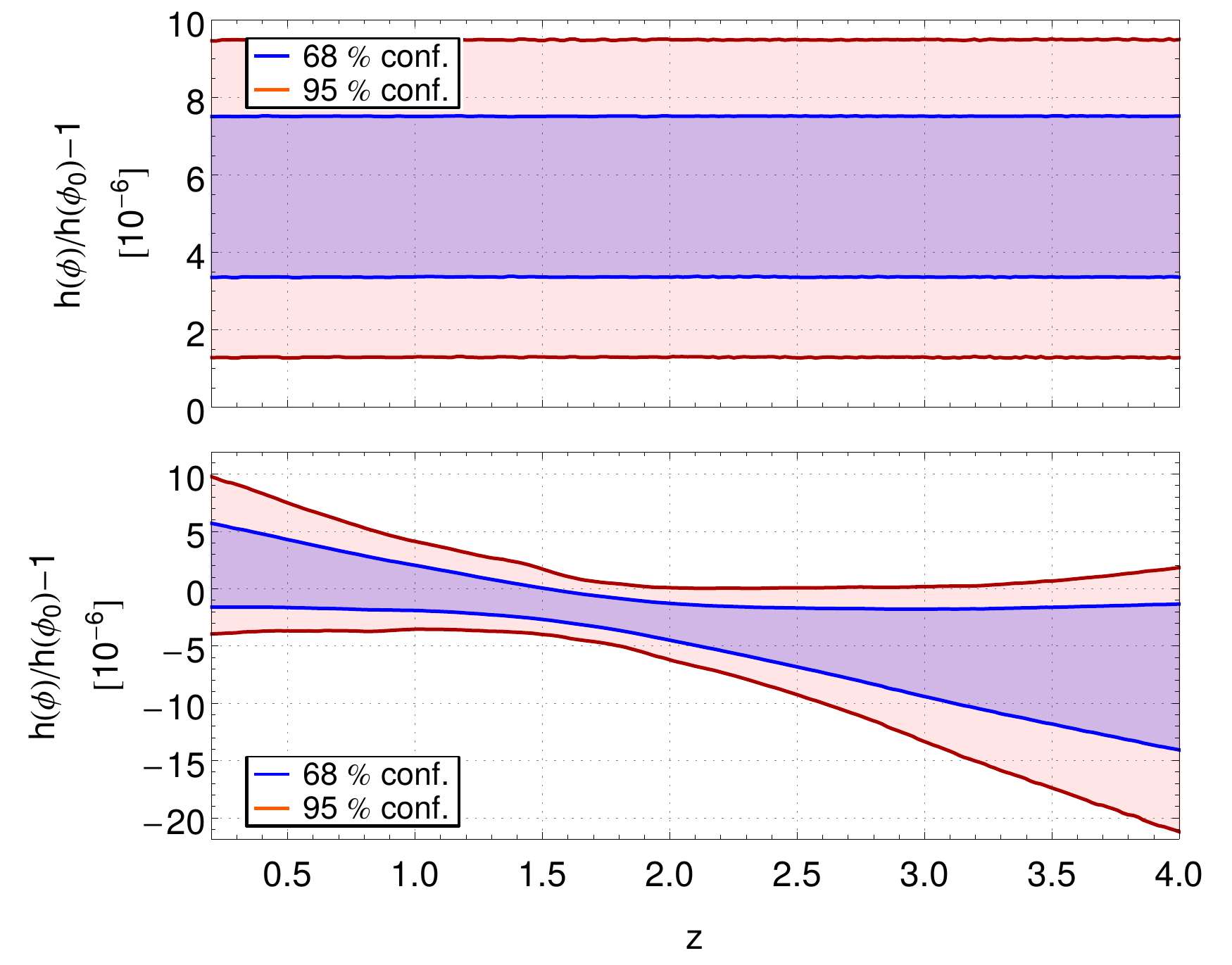}
\includegraphics[width=0.48\textwidth]{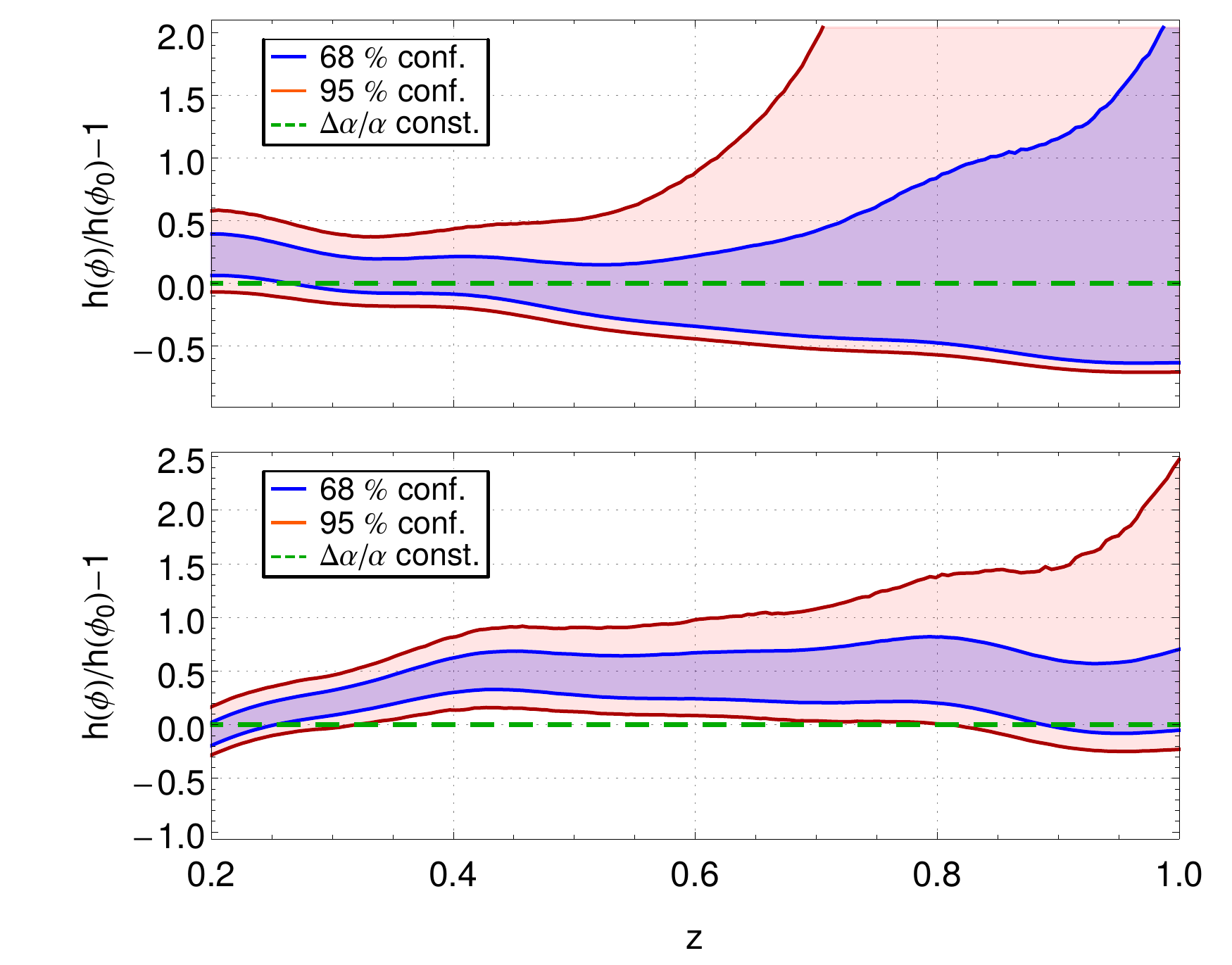}
\end{center}
\caption{Estimation of $h(\phi)/h(\phi_0)$ derived from constraints on $\Delta\alpha/\alpha$ (left) and on $\eta(z)$ (right) using Gaussian Processes. On top left: estimation done using Keck observations \cite{murphy:2003dz}. On bottom left: estimation done using VLT observations \cite{king:2012hb}. On right: estimations done from observations of luminosity distance \cite{suzuki:2012fk} and angular distance. On top right: the angular distances used are from \cite{de-filippis:2005pd}. On bottom right: the angular distances used are from \cite{bonamente:2006lq}.}
\label{fig:h}
\end{figure*}

First of all, we transform $\Delta\alpha/\alpha$ observations on estimations of the evolution of $h(\phi)/h(\phi_0)$ using the relation (\ref{eq:dalpha_eta}). We have used two sets of data:  values from 154 absorbers observed with the VLT (this data set can be found in \cite{king:2012hb}) and values from 128 absorbers observed at the Keck observatory (this dataset can be found in \cite{murphy:2003dz}). For both of these data sets, we have applied a GP, marginalized over the hyperparameters using a Markov chain Monte Carlo (MCMC) technique \cite{foreman-mackey:2013pb} which has produced a sample of the data at each reconstructed redshift $z$. These samples are then transformed into samples of $h(\phi)/h(\phi_0)$ using (\ref{eq:dalpha_eta}) and confidence intervals have been estimated. The left part of figure~\ref{fig:h} represents the confidence intervals obtained by using the two sets of data analyzed by using a GP.

The second type of observations we use is related to violations of the distance-duality relation $\eta(z)$. Indeed, the evolution of $h(\phi)/h(\phi_0)$ can also be estimated from $\eta(z)$ using (\ref{eq:etaM}). Two types of observations are needed in order to estimate $\eta$ (\ref{eq:eta}) : observations of luminosity distance  $D_L$ and of angular distance $D_A$. In this paper, we use the Supernovae Ia luminosity data from the Union 2.1 compilation \cite{suzuki:2012fk}. The luminosity distance is directly related to the distance modulus $\mu$ provided in \cite{suzuki:2012fk} by $D_L=10^{\mu/5 -5}$ where $D_L$ is expressed in Mpc. Regarding the angular distance, we use data from X-rays and Sunyaev-Zel'dovich observations of galaxy clusters. Two sets of data have been used \cite{de-filippis:2005pd,bonamente:2006lq} which provide $D_A^{\textrm{obs}}$ for different values of the redshift. As mentioned in \cite{uzan:2004lp} (see also \cite{cao:2011uq,holanda:2012yg}), if a violation of the distance-duality relation is considered, then the Sunyaev-Zel'dovich and X-rays observations measured $D_A^{\textrm{obs}}(z)=D_A(z)\eta^2(z)$. Therefore, an estimation of $\eta(z)$ from $D_A^{\textrm{obs}}(z)$ reads
\begin{equation}
 \eta(z)=\frac{D_A^{\textrm{obs}}(z)(1+z)^2}{D_L(z)}.
\end{equation}
The analysis procedure is similar to the one followed for the $\Delta\alpha/\alpha$ data. We have analyzed the $D_L$ and $D_A$ data with a GP, we have marginalized over the hyperparameters with a MCMC technique which has produced a sample of the data. This sample is then transformed into $\eta(z)$ and then to $h(\phi)/h(\phi_0)$ by using (\ref{eq:etaM}). From these, we can determine the confidence intervals that are represented on the right of figure~\ref{fig:h}. These estimations are 5 orders of magnitude larger than the ones obtained by using $\Delta\alpha/\alpha$ observations (that are represented by dashed lines). The estimation done using the set of data from \cite{bonamente:2006lq} shows a small deviation from the estimations done using a temporal variation of the fine structure constant between $z=0.4$ and $z=0.8$. If confirmed, this can be an indication of a violation of the coupling (\ref{eq:matterActionMod}) but at this stage, we believe it is the result of a lack of statistics. On the other hand, the estimation done using the set of data from \cite{de-filippis:2005pd} is in total agreement with the one from the variations of the fine structure constant. 

\begin{figure}[hbt]
\begin{center}
\includegraphics[width=0.49\textwidth]{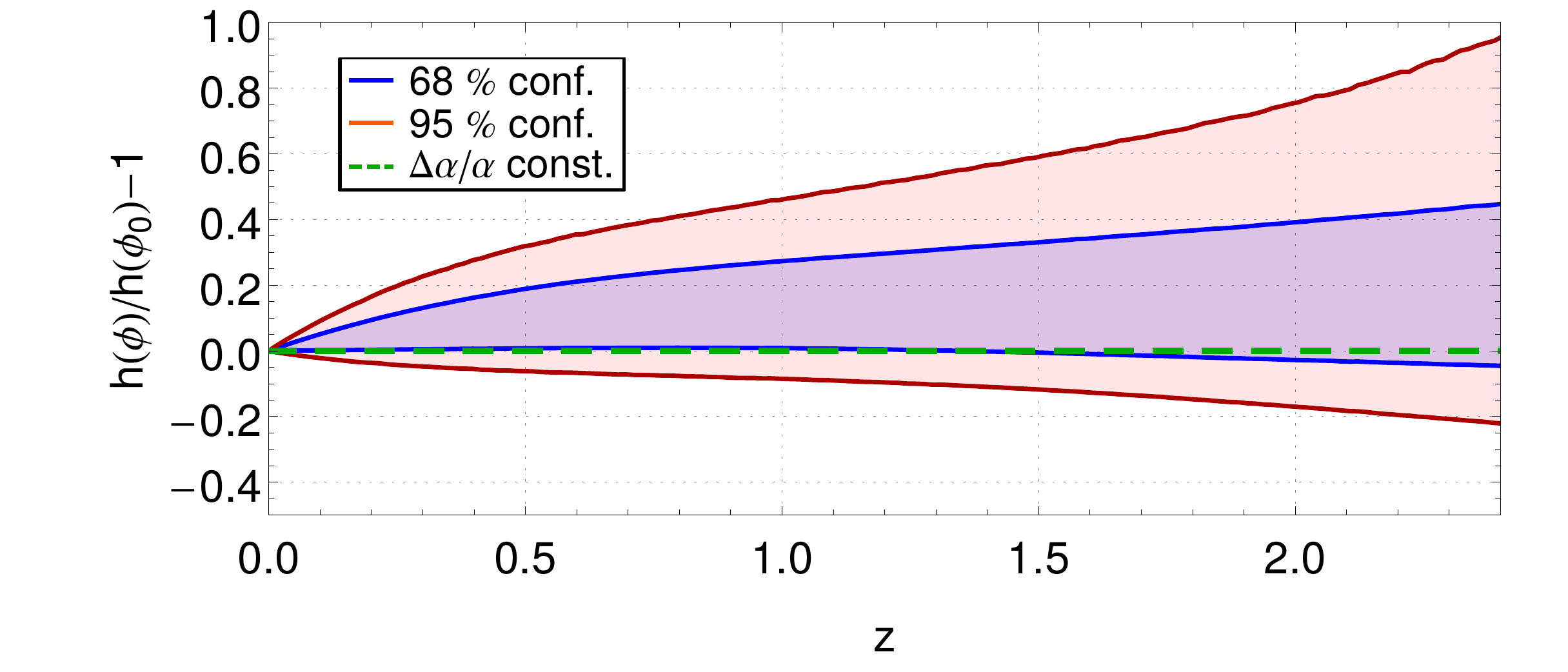}
\end{center}
\caption{Estimation of $h(\phi)/h(\phi_0)$ derived from constraints on $T_\textrm{CMB}(z)$ observations using GP.}
\label{fig:h_temp}
\end{figure}

Concerning the CMB temperature, we use data coming from Sunyaev-Zel'dovich observations at low redshifts \cite{luzzi:2009lr,saro:2014mn} and from observations of spectral lines at high redshift \cite{ge:1997yq,srianand:2000fk,molaro:2002uq,cui:2005fk,srianand:2008kx,carswell:2011rt,noterdaeme:2011vn,muller:2013ys,sato:2013zr}. In total, this represents 38 observations of the CMB temperature at redshift between 0 and 3. We also use the estimation of the current CMB temperature $T_0=2.725 \, K$ \cite{fixsen:2009bh}. The analysis procedure is similar to the ones used for the other observations. We have analyzed the temperature data using a GP, we have marginalized over the hyperparameters by using a MCMC technique which has provided a sample of the data. Then, we have transformed this sample into a sample of $h(\phi)/h(\phi_0)$ using the relation (\ref{eq:hphi}) and we have determined the confidence intervals. The figure~\ref{fig:h_temp} represents the estimation on the evolution of $h(\phi)/h(\phi_0)$ obtained from the CMB temperature observations. At low redshift, this estimation is roughly two times better than the one obtained by using the observations of the distances (see right of figure~\ref{fig:h}). In addition, the temperature measurements allow one to constrain the evolution of the scalar field at higher redshift. On the other hand, the constraints coming from the analysis of the observations of the temporal variation of the fine structure constant are 5 orders of magnitude better (see left of figure~\ref{fig:h}).

As a conclusion, all the data used seem to be consistent at the level of 10 \%. The observations of the variation of the fine structure constant are currently 5 orders of magnitude better than observations of the violation of the cosmic distance-duality and of the evolution of the CMB temperature. An improvement of the measurements of $\eta(z)$ and of the CMB temperature would be particularly useful in order to improve the test of the coupling (\ref{eq:matterActionMod}).

\subsection{Expected improvements with future experiments}
In this section, we will assess the improvements expected from future experiments focusing on the Square Kilometer Array (SKA) \cite{blake:2004xr} and on EUCLID \cite{amendola:2013lr}.

\begin{figure}[hbt]
\begin{center}
\includegraphics[width=0.49\textwidth]{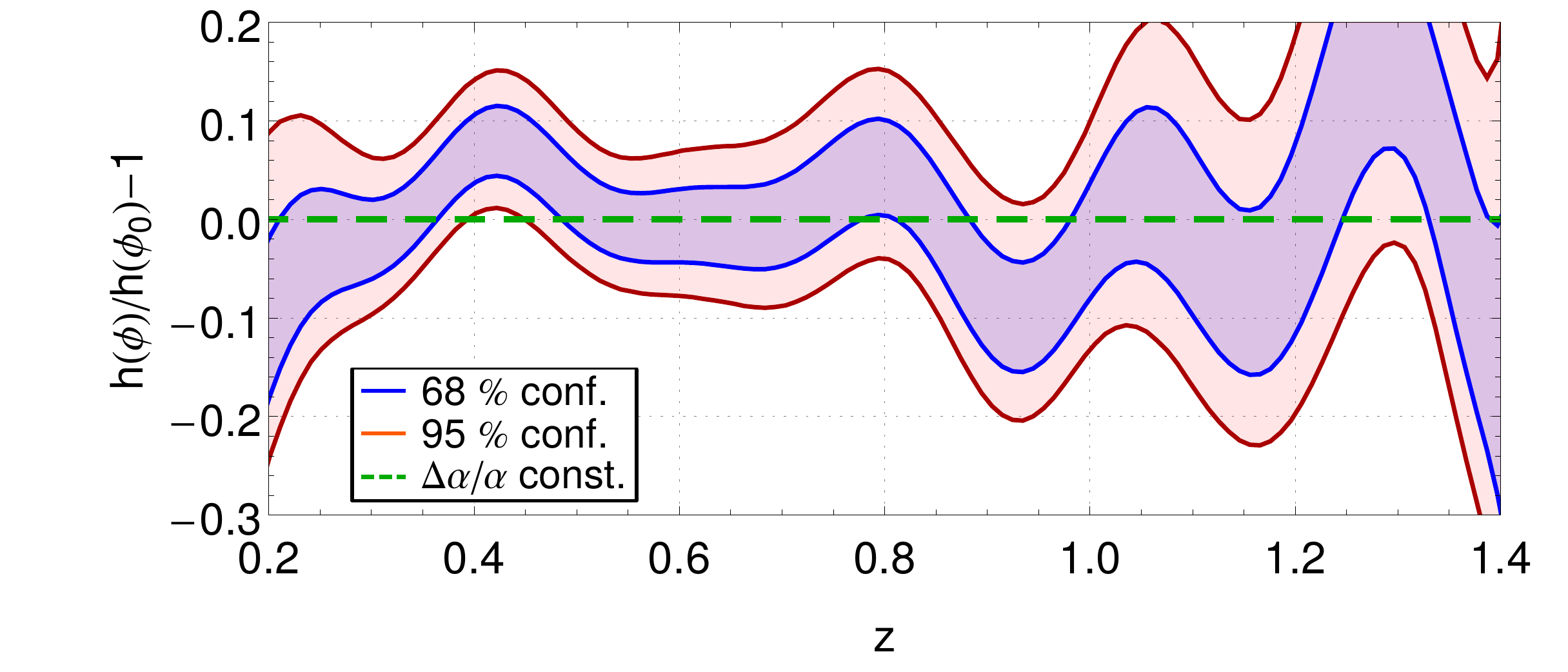}
\end{center}
\caption{Expected sensitivity by using Supernovae distance luminosity data \cite{suzuki:2012fk} and angular distance data from observations of Baryon Accoustic Oscillations (BAO) with SKA.}
\label{fig:ska}
\end{figure}

SKA will measure the angular distance $D_A(z)$ with Baryon Accoustic Oscillations (BAO) observations between $z=0.3$ and $z2$ \cite{bull:2014oz}. The expected accuracy of SKA is given in figure 6 of \cite{bull:2014oz} and is roughly 2 \% ($\sigma_{D_A}/D_A\sim 0.02$). Therefore, we simulated $D_A$ data from a standard scenario in GR and we reconstructed the estimation of $h(\phi)$ obtained assuming a 2\% relative accuracy on $D_A$. The figure~\ref{fig:ska} represents the obtained estimation. First of all, it is important to notice that the range of redshifts is larger than the one currently available (see on the right of figure~\ref{fig:h}). On figure~\ref{fig:ska}, we are now limited by the $D_L$ measurements that span $z=0-1.4$ only. Moreover, there is roughly one order of magnitude of improvement between current observations (right of figure~\ref{fig:h}) and what is expected with SKA. Nevertheless, this accuracy is still 4 orders of magnitude larger than the one obtained by using $\Delta\alpha/\alpha$ data (see left of figure~\ref{fig:h}).

EUCLID also expects to improve the constraint on the violation of the cosmic distance-duality relation thanks to measurements of the BAO \cite{amendola:2013lr}. In particular, EUCLID expect to constrain the parameter $\varepsilon$ from the parametrization (\ref{eq:parEps}) at a level better than $10^{-2}$ improving the current constraints by a factor of 5 (see table~\ref{tab:estdotalpha}).

Therefore, with these observations, we expect to improve the test of the coupling (\ref{eq:matterActionMod}) by one order of magnitude. Nevertheless, the observations of $\eta(z)$ will still remain 4 orders of magnitude less accurate than the one coming from the variations of the fine structure constant.

\section{Conclusion}
In this paper, we focused on cosmological signatures of modifications of gravity generated by a multiplicative coupling of a scalar field to the electromagnetic Lagrangian (\ref{eq:matterActionMod}). As mentioned in the introduction, this kind of coupling arises in various hypothetical alternative theories of gravity such as the low energy action of string theories \cite{damour:1994fk,damour:1994uq,gasperini:2002kx,minazzoli:2014pb}, in the context of axions \cite{peccei:1977lr,dine:1981pb,kaplan:1985nb}, of generalized chameleons \cite{brax:2004fk,brax:2007xi,weltman:2008jb,ahlers:2008mq} in Kaluza-Klein theories with additional compactified dimensions \cite{fujii:2003fi,overduin:1997wb}, in the Bekentein-Sandvik-Barrow-Magueijo theory of varying $\alpha$ \cite{bekenstein:1982zr,sandvik:2002ly,barrow:2012wd,barrow:2013rr}, in extended $f(R,\mathcal{L}_m)$ gravity \cite{harko:2013rz} or in the context of the pressuron \cite{minazzoli:2013fk}.

We have shown that this kind of coupling produces a temporal variation of the fine structure constant, a violation of the cosmic distance-duality relation, a modification of the evolution of the CMB temperature and CMB distortions. All these effects are intimately related to each other and to the cosmic evolution of the coupling $h(\phi)$ and we have derived relations between all these different observations.

Therefore, assuming that the coupling (\ref{eq:matterActionMod}) holds, which is the case for GR, for standard tensor-scalar theories with conformal coupling and for a large class of alternative theories of gravity, one can use the obtained relations to transform the constraints on one type of observation into constraints on another type of observation. We have used observations of variations of the fine structure constant to estimate the parameters of a violation of the distance-duality relation and the evolution of the CMB temperature. The obtained constraints are 5 orders of magnitude better than what is found in the literature but only hold for theories with a multiplicative coupling (\ref{eq:matterActionMod}) between the scalar field and the electromagnetic Lagrangian. These correspondences also allow  us to transform the constraint on the chemical potential of the CMB into a constraint on the variation of the fine structure constant between the CMB and the present epoch. 

On the other hand, comparing the different sets of observations allow one to test the coupling (\ref{eq:matterActionMod}). Indeed, a violation of the relation between $\Delta\alpha/\alpha$ and $\eta(z)$ or between $\Delta\alpha/\alpha$ and $T_{\textrm{CMB}}(z)$ would invalidate the multiplicative coupling independently of the form of the coupling function $h(\phi)$. To produce such a test, we transformed all the available observations into an estimation of the evolution of $h(\phi)$. This analysis was done by using Gaussian Processes. Then, we have compared the estimations provided by the different types of observations to detect any inconsistency that could result from a violation of the coupling (\ref{eq:matterActionMod}). We have shown that no inconsistency is currently detected. Moreover, observations from variations of the fine structure constant are currently 5 orders of magnitude better than observations from $\eta(z)$ and from the CMB temperature. The observations of the variations of the fine structure constant also predict a deviation of the distance-duality relation at the level of $10^{-6}$ in the case where the coupling (\ref{eq:matterActionMod}) is valid. For these reasons, it is particularly interesting to improve our constraints on $\eta(z)$ and on the evolution of the CMB temperature. One step to achieve this goal will be provided by planned future observations that will be done with the SKA or with EUCLID. In particular, the observations of the BAO will improve our measurements of the angular distance that will be reflected in an improvement on the constraint on $\eta(z)$ by one order of magnitude.

\begin{acknowledgments}
	 A. H. thanks A. Rivoldini for introducing him to the mystery of Bayesian inversion, H.L.~Bester for useful discussions about Gaussian Processes and A. F\"uzfa for interesting discussions about EUCLID. J. L. thanks P. Bull and P. Patel for help with the SKA accuracy forecasts. The authors thank A. Saro for sharing observational data on CMB temperature from \cite{saro:2014mn} and thank J. Barrow and J. Magueijo for interesting discussions on this topic.
\end{acknowledgments}

\bibliography{../../biblio_COPY}

\end{document}